\documentclass[prb,onecolumn,12pt]{revtex4-1}
\usepackage{epsfig}
\usepackage{amssymb}
\usepackage{threeparttable}
\usepackage{graphicx}
\usepackage{natbib}
\usepackage{longtable}
\usepackage{txfonts}
\usepackage{color}

\begin{document}
\title{High thermoelectric efficiency in monolayer PbI$_2$ from 300 K to 900 K}
\author{Bo Peng$^1$, Haodong Mei$^1$, Hao Zhang$^{1,*}$, Hezhu Shao$^2\dag$, Ke Xu$^1$,Gang Ni$^1$, Qingyuan Jin$^1$,,Costas M. Soukoulis$^{3,4}$ and Heyuan Zhu$^1$}
\affiliation{$^1$Key Laboratory for Information Science of Electromagnetic Waves (MOE) and Department of Optical Science and Engineering and Key Laboratory of Micro and Nano Photonic Structures (MOE), Fudan University, Shanghai 200433, China\\
$^2$Ningbo Institute of Materials Technology and Engineering, Chinese Academy of Sciences, Ningbo 315201, China\\
$^3$Department of Physics and Astronomy and Ames Laboratory, Iowa State University, Ames, Iowa 50011, USA\\
$^4$Institute of Electronic Structure and Laser (IESL), FORTH, 71110 Heraklion, Crete, Greece
}

\begin{abstract}
By using a first-principles approach, monolayer PbI$_2$ is found to have great potential in thermoelectric applications. The linear Boltzmann transport equation is applied to obtain the perturbation to the electron distribution by different scattering mechanisms. The mobility is mainly limited by the deformation-potential interaction with long-wavelength acoustic vibrations at low carrier concentrations. At high concentrations, ionized impurity scattering becomes stronger. The electrical conductivity and Seebeck coefficient are calculated accurately over various ranges of temperature and carrier concentration. The lattice thermal conductivity of PbI$_2$, 0.065 W/mK at 300 K, is the lowest among other 2D thermoelectric materials. Such ultralow thermal conductivity is attributed to large atomic mass, weak interatomic bonding, strong anharmonicity, and localized vibrations in which the vast majority of heat is trapped. These electrical and phonon transport properties enable high thermoelectric figure of merit over 1 for both p-type and n-type doping from 300 K to 900 K. A maximum $zT$ of 4.9 is achieved at 900 K with an electron concentration of 1.9$\times$10$^{12}$ cm$^{-2}$. Our work shows exceptionally good thermoelectric energy conversion efficiency in monolayer PbI$_2$, which can be integrated to the existing photovoltaic devices.
\end{abstract}

\maketitle

Organic-inorganic CH$_3$NH$_3$PbI$_3$ perovskite solar cells have emerged as a leading next-generation photovoltaic technology \cite{Kojima2009,Lee2012a,Park2013,Snaith2013,Kim2014,Habisreutinger2014,Green2014,Service2014,Hodes2014,Lin2015}. As the precursor material used to fabricate perovskite thin films \cite{Burschka2013,Jeng2013,Ng2015,Fu2015,Hill2015,Bae2016,Yi2016,Zhu2016,Raga2016,Noel2017}, lead iodide (PbI$_2$) leads to remarkable advances in efficiency due to the 6$s^2$ electronic configuration of Pb \cite{Ganose2016}. Encapsulated perovskite devices with excess PbI$_2$ exhibit good stability \cite{Liu2016c}. An excess of PbI$_2$ is beneficial to a better crystallization of the perovskite layer and improves the performance of perovskite solar cells \cite{Duong2016,Bi2016,Wang2016g,Kim2016}. After long exposures, CH$_3$NH$_3$PbI$_3$ eventually forms PbI$_2$ due to its instability in moist air \cite{Christians2015,Chen2014,Leguy2015,Yang2016c,Yang2015a}. According to this degradation process, waste PbI$_2$ at the end of its useful life can be recycled using an appropriate solvent \cite{Chen2014a,Babayigit2016}. Therefore, although CH$_3$NH$_3$PbI$_3$ perovskite has a drawback in the toxicity of lead \cite{Smith2014,Savory2016,Gupta2016}, the perovskite technology can be deployed in a completely safe way by recycling PbI$_2$.


After absorbing solar energy, the photo-induced carriers are generated in the CH$_3$NH$_3$PbI$_3$ region, while the PbI$_2$ passivation layers can prevent back recombination and facilitate charge separation \cite{Shih2017}. Besides the sunlight collected by the perovskite solar cells, a large fraction of solar energy is converted into heat in the form of phonons as well \cite{He2014a}. Such heat can be converted into electricity by thermoelectric materials when the temperature gradient is generated. Here we show for the first time that PbI$_2$ itself is a promising candidate for high-efficiency thermoelectric applications.



\begin{figure*}
\centering
\includegraphics[width=0.7\linewidth]{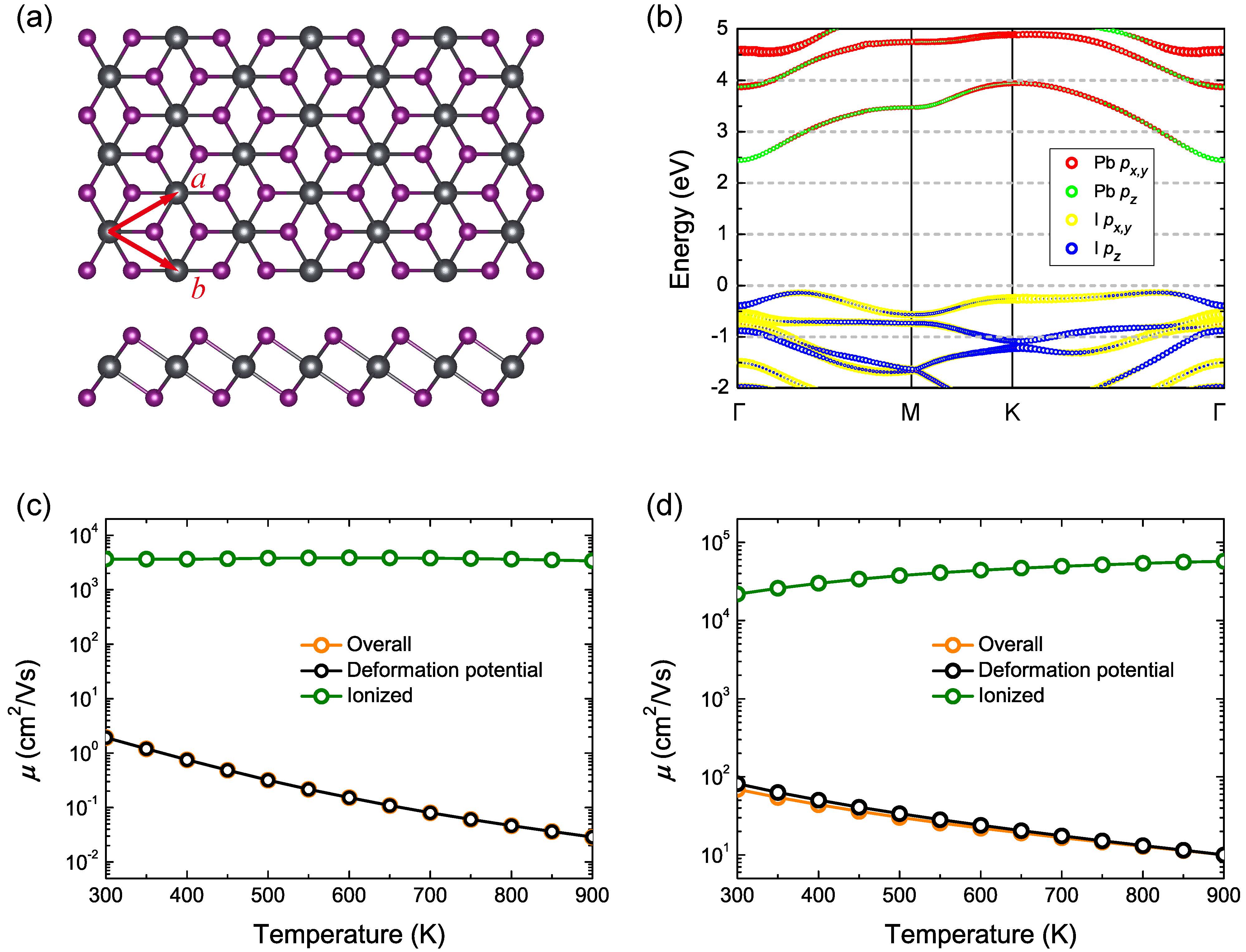}
\caption{(a) Crystal structure of monolayer PbI$_2$. (b) Projected orbital character of band structure of monolayer PbI$_2$ using the HSE functional with spin-orbit coupling. Calculated mobility from different scattering mechanisms of (c) p-type and (d) n-type PbI$_2$ with a carrier concentration of 1.9$\times 10^{9}$ cm$^{-2}$ at varying temperatures.}
\label{band}
\end{figure*}

The fabrication of PbI$_2$ nanostructures is being pursued with increasing interest in chemistry, physics, and material science \cite{Sandroff1986,Tan1987,Kasi2007,Baltog2009,Cabana2014,Zhong2016,Wang2016f,Zheng2016,Frisenda2017,Zhong2017,Wang2017b}. In this work, we focus on monolayer PbI$_2$ only, because of the following reasons: (1) Evidence for the reversible formation of monolayer PbI$_2$ has been found in 1987, which discovered that a monolayer precedes the production of bulk PbI$_2$ \cite{Tan1987}. By using cyclic voltammetry to study the electrocrystallisation of PbI$_2$, the monolayer is found to appear at an underpotential of -65 mV with respect to the reversible potential of crystalline PbI$_2$, and this can be explained by the reduction of surface tension which occurs when the solid electrode is covered by a monolayer. Under such laboratory conditions, monolayer PbI$_2$ is more stable than bulk PbI$_2$ by -13 kJ/$mol$ \cite{Tan1987}. (2) Bulk PbI$_2$ usually has a rough surface as well as lots of defects, which strongly reduce the carriers mobility \cite{Zhang2015j}, while in 2D PbI$_2$, low-density defects and ultrasmooth surface have been observed \cite{Zheng2016}. (3) Low dimensionality provides a effective conductive channel for carriers and reduces phonon thermal transport at the same time \cite{Arab2015}. (4) Most interestingly, it is possible to fabricate PbI$_2$ with other 2D materials by layer-engineering in photovoltaic and thermoelectric systems \cite{Zhou2015}. Thus monolayer PbI$_2$ is preferred for its transparency.

\begin{table*}
\centering
\caption{Calculated effective mass $m^*$, deformation-potential constant $E_d$, 2D modulus $C_{2D}$, overall carrier mobility $\mu$, mobility limited by deformation potential $\mu_{\textrm{de}}$ and ionized impurity scattering $\mu_{\textrm{ii}}$ of monolayer PbI$_2$ with a carrier concentration of 1.9$\times 10^{9}$ cm$^{-2}$ at 300 K.}
\begin{tabular}{ccccccccccc}
\hline
Carrier type & $m^*$ ($m_0$) & $E_d$ (eV) & $C_{2D}$ (J/m$^2$) & $\mu$ (cm$^2$/Vs) & $\mu_{\textrm{de}}$ (cm$^2$/Vs) & $\mu_{\textrm{ii}}$ ($\times 10 ^ {3}$ cm$^2$/Vs) \\
\hline
hole & 17.33 & -1.79 & 14.36 & 1.94 & 1.95 & 3.66 \\
electron & 0.73 & -4.41 & 14.39 & 69.52 & 81.66 & 21.73 \\
\hline
\end{tabular}
\label{mobility}
\end{table*}


Monolayer PbI$_2$ crystallizes in the space group $P3m1$. The optimized lattice constant of 4.66 \AA\ and height of 3.73 \AA\ are in good agreement with previous results \cite{Zhou2015,Lu2016a}. Each Pb atom is octahedrally surrounded by six I atoms, and I atoms themselves are hexagonally close packed, as shown in Figure~\ref{band}(a). 

We now turn to study the electronic structures and carrier mobility of monolayer PbI$_2$. The band structure is calculated using the HSE functional with spin-orbit coupling, as shown in Figure~\ref{band}(b). The calculated band gap is 2.57 eV, reproducing well the previous theoretical and experimental results \cite{Zhou2015,Zheng2016}. The conduction band minimum (CBM) is located at $\Gamma$ point, while the valence band maximum (VBM) shifts a little away from $\Gamma$ point. The VBM (CBM) bands mainly result from I-$5p$ (Pb-$6p$) states.

Both elastic (ionized impurity, piezoelectric, and deformation potential interaction) and inelastic (polar optical phonons) scattering mechanisms are taken into account in calculating the mobility $\mu$ \cite{Faghaninia2015}. The dielectric measurements show that for PbI$_2$, $\epsilon_0=\epsilon_{\infty}$=6.25 \cite{Dugan1967}. Thus the electron-polar optical phonon scattering rates, $\propto (1/\epsilon_{\infty}-1/\epsilon_0)$, are much lower than elastic scattering rates. It should be noticed that the $\epsilon_0$ and $\epsilon_{\infty}$ are measured on bulk PbI$_2$. For monolayer PbI$_2$, the low dimension reduces electronic screening, typically leading to smaller dielectric function. However, accurate estimation of dielectric function of monolayer PbI$_2$ requires many-body perturbation theory \cite{Peng2018c}. Here for simplicity we use the bulk value, and under such approximation, the electron-polar optical phonon scattering is negligible.

\begin{figure*}
\centering
\includegraphics[width=0.7\linewidth]{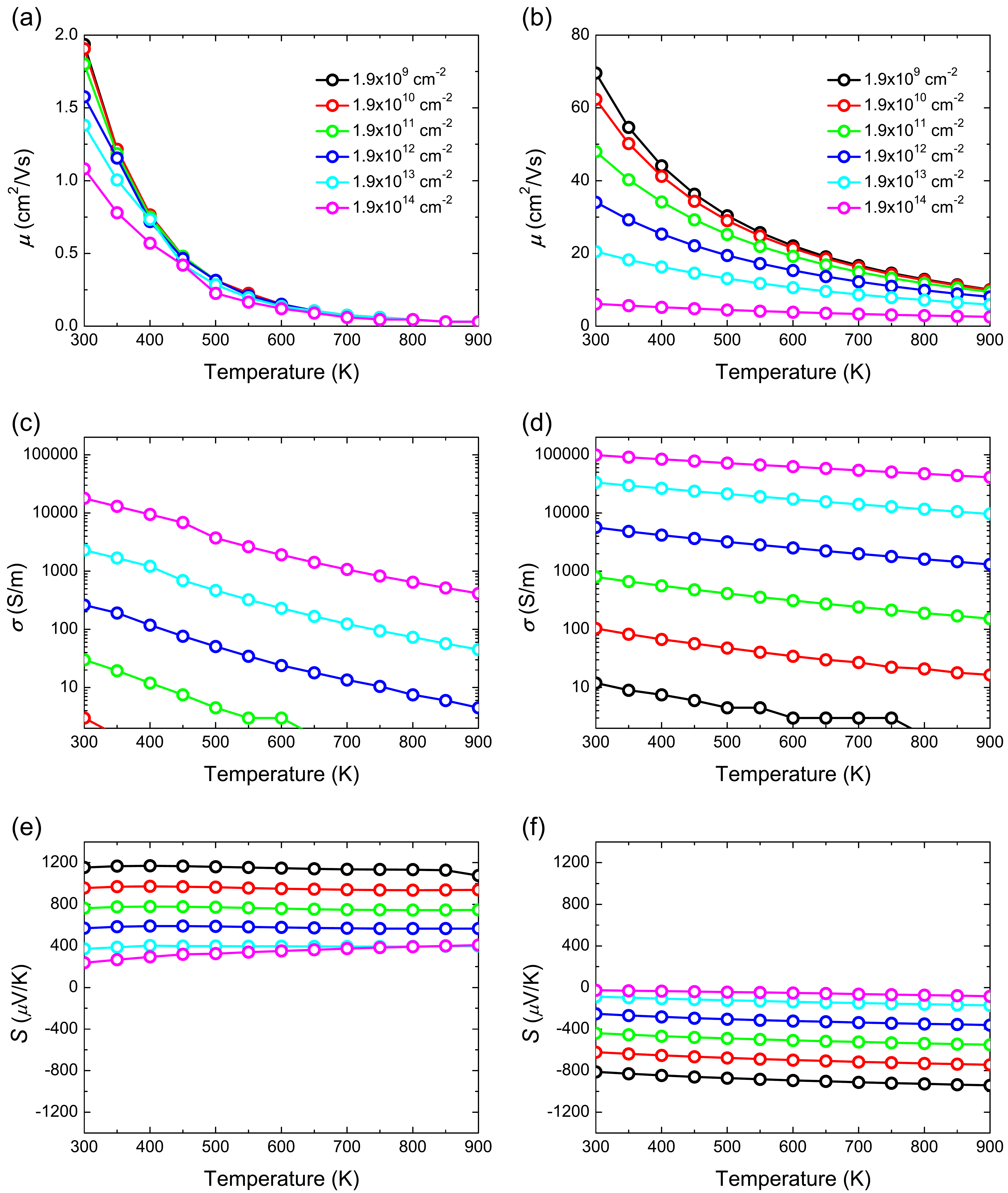}
\caption{Overall mobility $\mu$ with (a) p-type and (b) n-type doping, electrical conductivity $\sigma$ with (c) p-type and (d) n-type doping, and Seebeck coefficient $S$ with (e) p-type and (f) n-type doping as a function of temperature with different carrier concentrations for monolayer PbI$_2$.}
\label{electrical} 
\end{figure*}

The total elastic scattering rate can be calculated according to Matthiessen's rule \cite{Faghaninia2015}
\begin{equation}
\nu_{\textrm{el}} = \nu_{\textrm{de}} + \nu_{\textrm{ii}} + \nu_{\textrm{pe}},
\end{equation}
where $\nu_{\textrm{el}}$, $\nu_{\textrm{de}}$, $\nu_{\textrm{ii}}$, and $\nu_{\textrm{pe}}$ stand for elastic, deformation potential, ionized impurity, and piezoelectric scattering rates, respectively. Table~\ref{mobility} lists the related \textit{ab initio} parameters for calculating the elastic scattering rates using the single-band approximation, as well as the overall mobility and the mobility limited by deformation potential and ionized impurity scattering with a carrier concentration of 1.9$\times 10^{9}$ cm$^{-2}$ at 300 K.

The calculated mobilities from different scattering mechanisms are shown in Figure~\ref{band}(c) and (d) for both p-type and n-type doping. Although it might be very difficult for n-type doping as monolayer PbI$_2$ has a large band gap, it can be realized by electrostatic gating \cite{Li2016c}. By changing the gate voltage, the injected charge can be tuned \cite{Sun2018}. With a zero piezoelectric coefficient $e_{11}$, monolayer PbI$_2$ does not exhibit pronounced piezoelectricity. This is because monolayer PbI$_2$ has inversion center, as been observed in the 1T structure of SnS$_2$ \cite{Blonsky2015}. For comparison, monolayer MoS$_2$, which has 2H structure and hence breaks inversion symmetry, shows a piezoelectric coefficient of 362 pC/m \cite{Blonsky2015}. Thus the mobility cannot be limited by piezoelectric scattering.

\begin{figure*}
\centering
\includegraphics[width=0.7\linewidth]{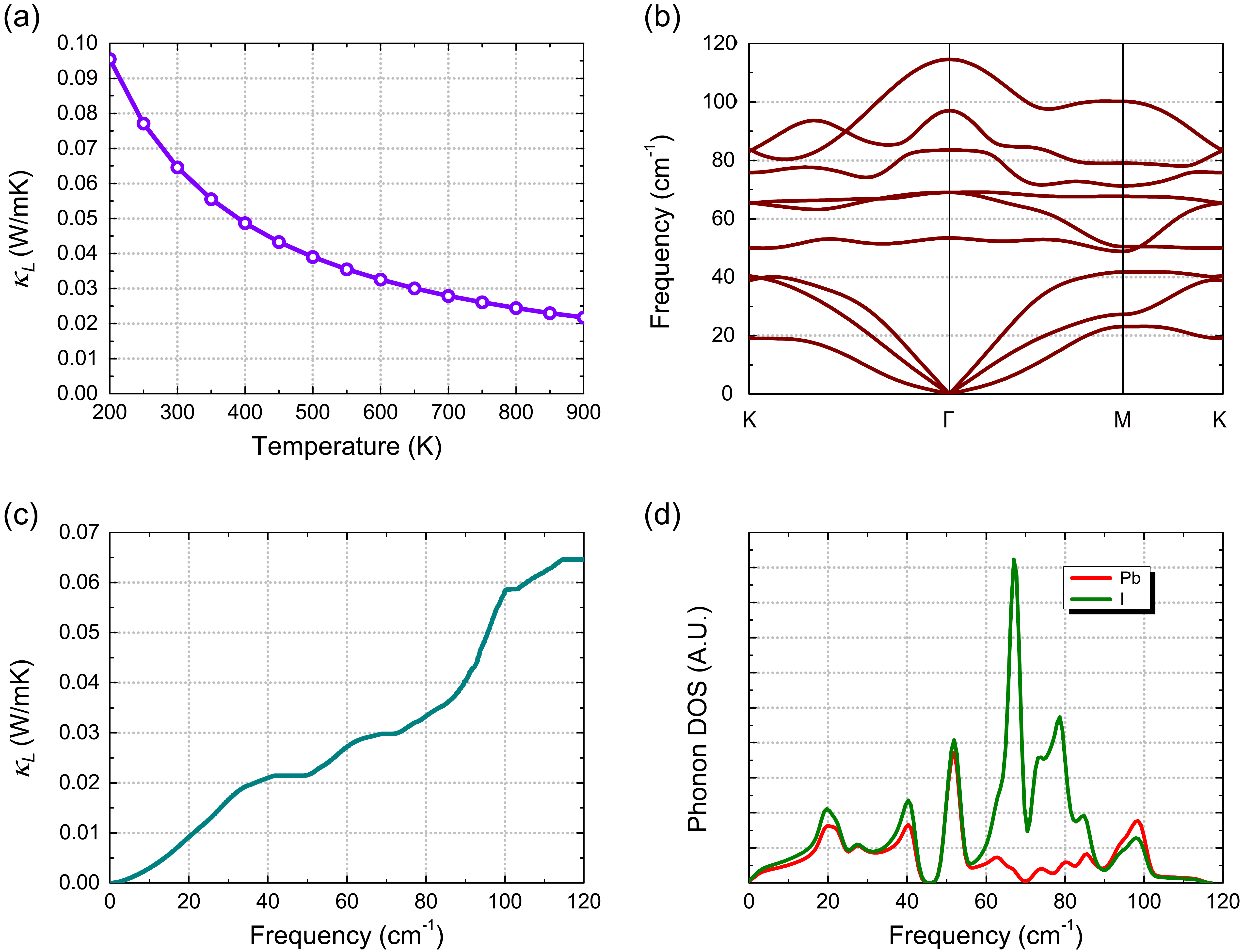}
\caption{(a) Lattice thermal conductivity as a function of temperature, (b) phonon dispersion, (c) cumulative thermal conductivity as a function of phonon frequency and (d) projected phonon density of states for monolayer PbI$_2$.}
\label{phonon} 
\end{figure*}

For both p-type and n-type PbI$_2$ with a carrier concentration of 1.9$\times 10^{9}$ cm$^{-2}$, the mobility at 300 K is mainly limited by deformation potential interaction with acoustic phonons. At low carrier concentrations, the wavelength of thermally activated carriers is much larger than the lattice constant. Therefore the carrier mobility is determined by acoustic vibrational modes \cite{Bardeen1950,Xue2016}. In deformation potential interaction, due to much smaller effective mass at the CBM, the electron mobility is much larger than hole mobility.

To determine the transport properties, we use the rigid band approximation, in which the electronic structure is unchanged with doping and only the Fermi level is shifted appropriately. For impurity doping, ionized impurities become scatterer centers and their concentration can be calculated at a given carrier concentration by iteratively solving the charge balance equation \cite{Faghaninia2015}. Figure~\ref{electrical}(a) and (b) show the calculated $\mu$. With increasing carrier concentration, the ionized impurity scattering becomes stronger, which further reduces the carrier mobility.

Once $\mu$ is calculated, the electrical conductivity can be obtained at a given carrier concentration (assuming that the carrier concentration remains constant at different temperatures). As shown in Figure~\ref{electrical}(c) and (d), the $\sigma$ increases with increasing carrier concentration and decreases with increasing temperature.

The Seebeck coefficient $S$ measures the electrical potential difference created from a temperature gradient. As shown in Figure~\ref{electrical}(e) and (f), the absolute values of $S$ are larger than 400 $\mu$V/K at low carrier concentrations over a large range of temperature, which is larger than those of antimonene \cite{Peng2018a}. At a hole concentration of 1.9$\times 10^{9}$ cm$^{-2}$, the $S$ reaches nearly 1200 $\mu$V/K. The absolute values of $S$ decrease with increasing carrier concentration. It should be noticed that, although the constant relaxation time approximation correctly predicts the trend in Seebeck coefficient with varying carrier concentration \cite{Madsen2006}, the predicted values are far from the experimental results because the treatment of relaxation time as a single constant affects both $\sigma$ and $S$ when integrated over energy.

\begin{figure*}
\centering
\includegraphics[width=0.7\linewidth]{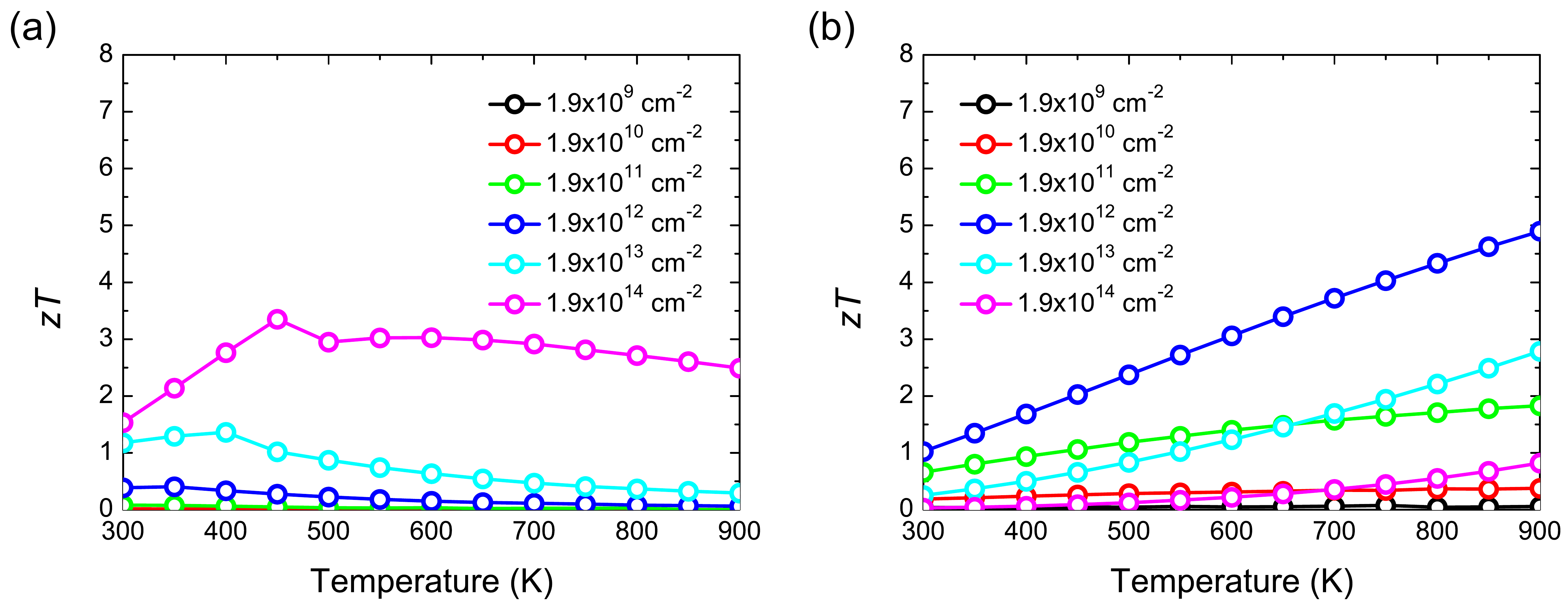}
\caption{Thermoelectric $zT$ of PbI$_2$ along $a$ and $b$ directions as a function of the carrier concentration at different temperatures for p-type and n-type doping.}
\label{zt} 
\end{figure*}

The key and remarkable feature of monolayer PbI$_2$ is ultralow lattice thermal conductivity $\kappa_L$, which ranges from 0.096 W/mK at 200 K to 0.022 W/mK at 900 K. This extraordinarily low $\kappa_L$ is much lower than other 2D thermoelectric material \cite{Wang2015d,Wang2013a,Sevincli2014,Tan2015b,Lv2016,Peng2018a}. To quantitatively understand the origin of ultralow $\kappa_L$ in monolayer PbI$_2$, we compare the results using the Boltzmann transport equation for phonons in Figure~\ref{phonon}(a) with the Slack model \cite{Slack1973321}.

According to the Slack model, the $\kappa_L$ is given by \cite{Yang2016}
\begin{equation}
\kappa_L=A\frac{\bar{m} \theta_D^3\delta}{\gamma^2 n^{2/3}T}
\end{equation}
where $\bar{m}$ is the average mass, $\theta_D$ is the Debye temperature, and $\delta^2$ is the area per atom, $\gamma$ is the Gr\"uneisen parameters, $A$ is a $\gamma$ related parameter \cite{Julian1965}, and $n$ is the number of atoms in the unit cell. 

The calculated $\kappa_L$ of 0.070 W/mK is close to that using the Boltzmann transport equation for phonons (0.065 W/mK). The Slack model attempts to normalize the effect of crystal structure, atomic mass, interatomic bonding, and anharmonicity. Monolayer PbI$_2$ has a large atomic mass of 461 amu, while a low Young's modulus $Y_{2D}$ of 13.61 N/m indicates weak interatomic bonding. Large atomic mass and weak interatomic bonding lead to a low Debye temperature of 123 K. In addition, as shown in Figure S1 in the Supporting Information, the mode Gr\"uneisen parameters of PbI$_2$ are comparable to those of state-of-the-art thermoelectric materials such as PbTe and SnTe \cite{Lee2014}. Strong anharmonicity indicates strong three-phonon scattering strength \cite{Peng2016g}, which is the dominant scattering mechanism in phonon transport of monolayer PbI$_2$. Therefore, although monolayer PbI$_2$ and monolayer ZrS$_2$ have same crystal structure, due to large atomic mass, weak interatomic bonding and strong anharmonicity, the $\kappa_L$ of PbI$_2$ is much lower than that of ZrS$_2$ (3.29 W/mK at 300 K \cite{Lv2016}).



To further understand the origin of ultralow thermal conductivity in PbI$_2$, we examine the phonon vibrational properties. The calculated phonon dispersion is shown in Figure~\ref{phonon}(b). The highest phonon frequency of monolayer PbI$_2$ is 114.6 cm$^{-1}$, which is lower than that of state-of-the-art thermoelectric material PbTe \cite{Tian2012} and Bi$_2$Te$_3$ \cite{Qiu2009}, resulting in low phonon group velocity. Phonons with small group velocities are not effective carriers of heat \cite{Feldman1993}. 

In addition, as shown in Figure~\ref{phonon}(d), the flattened dispersions, corresponding to the peaks of phonon density of states from 40 cm$^{-1}$ to 80 cm$^{-1}$, imply localized phonon vibrations. It is well known that the flat modes tend to increase the number of three-phonon scattering channels \cite{Li2015a,Peng2016g}. Besides increased scattering channels, localized phonon states also result in reductions in the group velocities. As a result, the vast majority of heat is trapped in flat, low velocity modes, as shown in Figure~\ref{phonon}(c). Therefore phonons with frequencies from 40 cm$^{-1}$ to 80 cm$^{-1}$ contributes far less than those below 40 cm$^{-1}$ or beyond 80 cm$^{-1}$.



Generally, high thermoelectric performance is found in materials with high Seebeck coefficient $S$, high electrical conductivity $\sigma$, and low thermal conductivity $\kappa$, and the efficiency is determined by the dimensionless figure of merit $zT$ \cite{Zeier2016,Yang2016}
\begin{equation}
zT=\frac{\sigma S^2 T}{\kappa},
\end{equation}
where $\kappa=\kappa_e+\kappa_L$ is thermal conductivity consisting of electronic and lattice contributions. The electronic thermal conductivity $\kappa_e$ relates to the electrical conductivity $\sigma$ via the Wiedemann-Franz law $\kappa_e = L \sigma T$, where $L$ is the Lorenz number \cite{Peng2016}. Combining electrical and phonon transport properties, the thermoelectric figure of merit $zT$ in monolayer PbI$_2$ is evaluated in Figure~\ref{zt}. 

The highest $zT$ reaches 4.9 at 900 K with an electron concentration of 1.9$\times$10$^{12}$ cm$^{-2}$. Due to ultralarge Seebeck coefficient and ultralow thermal conductivity, high $zT$ values over 1 are achieved in a wide temperature range from 300 K to 900 K, while most thermoelectric materials appear promising only at mid or high temperatures \cite{Zhao2014,Lin2016,Ying2016}. Even for BiCu$_{0.7}$Ag$_{0.3}$SeO, the maximum $zT$ at 300 K is only 0.07, while PbI$_2$ reaches a $zT$ of 1, distinguishing itself for low temperature thermoelectric applications. Moreover, PbI$_2$ has been fabricated with perovskite thin film. It is tempting to build an additional PbI$_2$-based thermoelectric device to harvest the heat produced by perovskite-based photovoltaic devices.



To conclude, we show that PbI$_2$ is a promising thermoelectric material in a wide temperature range. By considering both the elastic and inelastic scattering mechanisms, highly accurate electrical transport properties are calculated using the linear Boltzmann transport equation. The acoustic-phonon limited mobilities at 300 K are 1.94 and 69.52 cm$^2$/Vs for p-type and n-type doping at a carrier concentration of 1.9$\times 10^{9}$ cm$^{-2}$, respectively. Large Seebeck coefficients are observed over large ranges of temperature and carrier concentration. Monolayer PbI$_2$ exibits an ultralow lattice thermal conductivity of 0.065 W/mK at 300 K. The origin of the intrinsically low lattice thermal conductivity is due to large atomic mass, weak interatomic bonding and strong anharmonicity. Lattice dynamics calculations show that weak bonding interactions lead to localized vibrations, and consequently the vast majority of heat is trapped in these modes due to increased scattering channels and reduced group velocities. By integrating all these features, both p-type and n-type monolayer PbI$_2$ exhibits a $zT$ over 1 from 300 K to 900 K at certain carrier concentrations, enabling flexible applications in thermoelectrics. In particular, we achieve a maximum $zT$ of 4.9 at 900 K with an electron concentration of 1.9$\times$10$^{12}$ cm$^{-2}$. Considering lead iodide perovskites are widely employed in solar cells recently, it is possible to fabricate PbI$_2$ with perovskite thin film in a hybrid thermoelectric and photovoltaic system, which may open up a path to a sustainable energy future. Experimental investigations are called for to verify our predictions and realize such devices in an industrially feasible way.


\section*{Methods}

First principles calculations are performed using the Vienna \textit{ab-initio} simulation package (VASP) based on density functional theory (DFT) \cite{Kresse1996}. The generalized gradient approximation (GGA) in the Perdew-Burke-Ernzerhof (PBE) parametrization for the exchange-correlation functional is used. A plane-wave basis set is employed with kinetic energy cutoff of 600 eV. We use the projector-augmented-wave (PAW) potential with 5\textit{d} electrons of Pb described as valence. A 15$\times$15$\times$1 \textbf{k}-mesh is used during structural relaxation for the unit cell until the energy differences are converged within 10$^{-8}$ eV, with a Hellman-Feynman force convergence threshold of 10$^{-6}$ eV/\AA. We maintain the interlayer vacuum spacing larger than 15 \AA\ to eliminate interactions between adjacent layers.

Hybrid functional methods based on the Heyd-Scuseria-Ernzerhof method are also adopted \cite{HSE1,HSE2,HSE3} with a 11$\times$11$\times$1 \textbf{k}-mesh. The Wannier functions are generated for generic band interpolation with a 31$\times$31$\times$1 \textbf{k}-mesh \cite{Mostofi2014}. The electrical transport properties are in-plane isotropic and can be calculated using the Boltzmann transport equation \cite{Faghaninia2015}. The band structure, density of state, electron group velocity, valence and conduction band wave function admixture, deformation-potential constant, 2D modulus, and polar optical phonon frequency are used in calculating the mobility. Based on the deformation potential theory in 2D materials \cite{Bruzzone2011,Xi2012,Cai2014a,Qiao2014,Dai2015}, we calculate the 2D elastic modulus and the deformation potential constant from the total energy
and the positions of CBM and VBM with respect to the lattice dilation up to 1.5\%. In calculating the piezoelectric constant \cite{Rode1975}, the elastic tensor including ionic relaxations is calculated using the finite differences method \cite{LePage2002,Wu2005,Peng2017a}, and the piezoelectric tensor is calculated using density functional perturbation theory (DFPT) \cite{Wu2005}. Because the electrons are confined in 2D, the in-plane mobility is 3/2 the average mobility for isotropic bulk materials. After explicitly considering the elastic and inelastic scattering, the electrical conductivity and Seebeck coefficient are calculated over large ranges of temperature and carrier concentration. The 2D carrier concentration is renormalized by the vacuum space between the 2D layers. The constant relaxation time approximation implemented in the BoltzTraP code is also used for comparison \cite{Madsen2006,Yang2008}, which gives similar trend in $\sigma$ and $S$.

The phonon transport properties of PbI$_2$ are calculated using the Boltzmann transport equation for phonons \cite{Omini1996,Li2012a,Li2012,Li2013a,ShengBTE}. The in-plane $\kappa$ is isotropic and can be calculated iteratively using the ShengBTE code as a sum of contribution of all the phonon modes \cite{Omini1996,Li2012a,Li2012,Li2013a,ShengBTE}. The harmonic interatomic force constants (IFCs) are obtained by DFPT using a 5$\times$5$\times$1 supercell with 5$\times$5$\times$1 \textbf{q}-mesh \cite{DFPT}. The Debye temperature is calculated from the average sound velocity \cite{Peng2016f,Peng2017}. The anharmonic IFCs are calculated using a supercell-based, finite-difference method \cite{Li2012}, and a 3$\times$3$\times$1 supercell with 5$\times$5$\times$1 \textbf{q}-mesh is used. We include the interactions with the eighth nearest-neighbor atoms (8.9 \AA). We use the nominal layer thicknesses $h$=6.98 \AA\ for PbI$_2$, corresponding to the interlayer distance of bulk PbI$_2$ \cite{Kasi2007}. The convergence of thermal conductivity with respect to $\textbf{q}$ points is tested in our calculation. A discretizationa of the Brillouin zone (BZ) into a $\Gamma$-centered regular grid of 91$\times$91$\times$1 $\textbf{q}$ points is introduced with scale parameter for broadening chosen as 1. It should be noticed that the monolayer and thin film thermal conductivities show different behaviours. The accurate estimation of thin film phonon transport requires calculating the effective in-plane thermal conductivity \cite{Chen2019}, which is not the topic of this manuscript.

\subsection*{Experimental section}

See in the supplementary material.

\section*{Acknowledgement}

This work is supported by the National Natural Science Foundation of China under Grants No. 11374063 and 11404348, and the National Basic Research Program of China (973 Program) under Grant No. 2013CBA01505. Work at Ames Laboratory was partially supported by the U.S.Department of Energy, Office of Basic Energy Science, Division of Materials Science and Engineering (Ames Laboratory is operated for the U.S. Department of Energy by Iowa State University under Contract No. DE-AC02-07CH11358). The European Research Council under ERC Advanced Grant No. 320081 (PHOTOMETA) supported work at FORTH.

\section*{References}

%

\end{document}